\RequirePackage{fix-cm}
\documentclass[smallextended]{svjour3}       % onecolumn (second format)
\smartqed  % flush right qed marks, e.g. at end of proof
\usepackage{graphicx}
\usepackage{tikz}
\usepackage[multi-part-units=single]{siunitx}
\usepackage{multirow}
\usepackage{upgreek}
\usepackage{fixltx2e}
\usepackage[misc]{ifsym}
\usepackage{float}
\usepackage[font=small,labelfont=bf,tableposition=top]{caption}
\usepackage{mathtools}
\usepackage{xcolor}
\usepackage{amsmath}
\usepackage{amssymb}
\begin{document}

\title{Spatio-Temporal Deep Learning Methods for Motion Estimation Using 4D OCT Image Data
%\thanks{Grants or other notes
%about the article that should go on the front page should be
%placed here. General acknowledgments should be placed at the end of the article.}
}
%\subtitle{Do you have a subtitle?\\ If so, write it here}

\titlerunning{Deep Learning Methods for Motion Estimation Using 4D OCT Image Data}        % if too long for running head

\author{Marcel Bengs$^{1}$$^{*}$ \and Nils Gessert$^{1}$$^{*}$ \and Matthias Schlüter$^{1}$ \and Alexander Schlaefer$^{1}$}

\authorrunning{Bengs et al.}

\institute{\Letter \quad Marcel Bengs, \email{marcel.bengs@tuhh.de}, Tel.: +49 (0)40 42878 3389 \\ $^{*}$ Authors contributed equally  \\ $^1$ Institute of Medical Technology, Hamburg University of Technology, Hamburg, Germany}

\date{Preprint. Accepted for publication in IJCARS.}
% The correct dates will be entered by the editor

\maketitle

% original article 3000–5000 words 
\begin{abstract}

\textit{Purpose} Localizing structures and estimating the motion of a
specific target region are common problems for navigation during surgical
interventions. Optical coherence tomography (OCT) is an imaging
modality with a high spatial and temporal resolution that has been
used for intraoperative imaging and also for motion estimation, for
example, in the context of ophthalmic surgery or cochleostomy.
Recently, motion estimation between a template and a moving OCT image
has been studied with deep learning methods to overcome the
shortcomings of conventional, feature-based methods.

\textit{Methods}  We investigate whether using a temporal stream of OCT image volumes can improve deep learning-based motion estimation performance. For this
purpose, we design and evaluate several 3D and 4D deep learning
methods and we propose a new deep learning approach. Also, we propose a temporal regularization strategy at the model output.

\textit{Results} Using a tissue dataset without additional markers, our deep learning methods using 4D data outperform previous approaches. The best performing 4D architecture achieves an correlation coefficient (aCC) of 98.58\% compared to 85.0\% of a previous 3D deep learning method. Also, our temporal regularization strategy at the output further improves 4D model performance to an aCC of 99.06\%. In particular, our 4D method works well for larger motion and is robust towards image rotations and motion distortions.

\textit{Conclusions} We propose 4D spatio-temporal deep learning for
OCT-based motion estimation. On a tissue dataset, we find that using 4D information for the model input improves performance while
maintaining reasonable inference times. Our regularization strategy
demonstrates that additional temporal information is also beneficial
at the model output.

\keywords{4D Deep Learning \and Optical Coherence Tomography \and Motion Estimation \and Regularization}

\end{abstract}

\section{Introduction}
\label{intro}

%Motivate OCT
Optical coherence tomography (OCT) is an image modality that is based on optical backscattering of light and allows for volumetric imaging with a high spatial and temporal resolution \cite{siddiqui2018high}. The imaging modality has been integrated into intraoperative microscopes \cite{lankenau2007combining} with applications to neurosurgery \cite{finke2012automatic} or ophthalmic surgery \cite{ehlers2014integrative}. Moreover, OCT has been used for monitoring laser cochleostomy \cite{pau2008imaging}. 

%Small FOV
While OCT offers a high spatial and temporal resolution, its field of view (FOV) is typically limited to a few millimeters or centimeters \cite{kraus2012motion}. Therefore, during intraoperative imaging, the current region of interest (ROI)  can be lost quickly due to tissue or surgical tool movement, which requires constant tracking of the ROI and corresponding adjustment of the FOV. Performing the adjustment manually can disrupt the surgical workflow which is why automated motion compensation would be desirable.
In addition to that, some surgical procedures such as laser cochleostomy also require adjustment of a surigcal tool in case patient motion occurs \cite{zhang2014optical}.
Due to the small scale of the cochlea structure, accurate adjustment is critical to avoid damaging surrounding tissue \cite{bergmeier2017workflow}. Both motion compensation for the adjustment of the OCT's FOV and the adjustment of surgical tools require accurate motion estimation.

One approach is to use an external tracking system for motion estimation. For example, Vienola et al. used this approach with a scanning laser ophthalmoscope for motion estimation in the context of FOV adjustment \cite{vienola2012real}. Also, external tracking systems have been used in the context of cochleostomy \cite{eilers2009navigated,du2013robustness}. Alternatively, the OCT images can be used directly for motion estimation as OCT already offers a high spatial resolution. For example, Irsch et al. estimated the tissue surface distance from A-scans for axial FOV adjustment \cite{irsch2018motion}. Also, Laves et al. used conventional features such as SIFT \cite{lowe1999object} and SURF \cite{bay2006surf} with 2D maximum intensity projects for motion estimation in the context of volume of interests stabilization with OCT \cite{laves2017feature}. Another approach for high-speed OCT tracking relied on phase correlation for fast motion estimation from OCT images \cite{schluter2019feasibility}.  
These approaches rely on hand-crafted features which can be error-prone and the overall motion estimation accuracy is often limited \cite{laves2019deep}. Therefore, deep learning methods have been proposed for motion estimation from OCT data. For example, Gessert et al. proposed using 3D convolutional neural networks (CNNs) for estimating a marker's pose from single 3D OCT volumes \cite{gessert2018deep}. For estimating the motion between two subsequent OCT scans, Laves et al. adopted a deep learning-based optical flow method \cite{ilg2017flownet} using 2.5D OCT projections \cite{laves2019deep}. Similarly, Gessert et al. proposed a deep learning approach for motion estimation where the parameters for a motion compensation system are directly learned from 3D OCT volumes by a deep learning model \cite{gessert2019two}. 

So far, deep learning-based motion estimation with OCT relied on an initial template volume and a moving image, following the concept of registration-based motion estimation, for example, using phase correlation \cite{schluter2019feasibility}. This can be problematic if motion between the original template and the current state is very large as the overlap between the images becomes small. Modern OCT systems could overcome this problem by acquiring entire sequences of OCT volumes, following the motion trajectory, as very high acquisition rates have been achieved \cite{wang2016heartbeat}. Therefore, more information can be made available between an initial state and the current state which could be useful for motion estimation. While deep learning approaches using two images could follow the trajectory with pair-wise comparisons, we hypothesize that processing an entire sequence of OCT volumes at once might provide more consistence and improved motion estimation performance.

\begin{figure}
\centering
\includegraphics[width=1.0\textwidth]{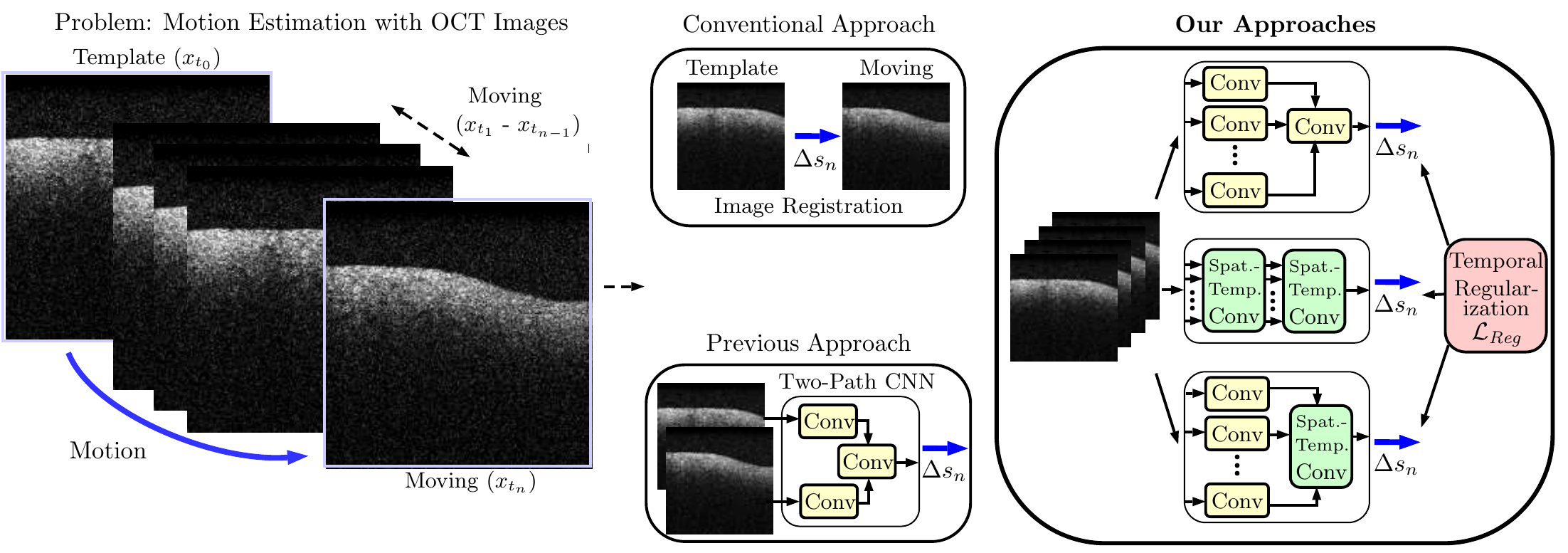}
\caption{Our approach for motion estimation in comparison to previous methods. The approach is illustrated for 2D OCT images for simplicity. Note, we perform all experiments with 3D volumetric OCT images and thus 4D spatio-temporal data.}
\label{fig:Motivation_Image}
\end{figure}

In this paper we compare several deep learning methods and investigate whether using 4D spatio-temporal OCT data can improve deep learning-based motion estimation performance, see Figure~\ref{fig:Motivation_Image}. Using 4D data with deep learning methods is challenging in terms of architecture design due to the immense computational and memory requirements of high-dimensional data processing. In general, there are only few approaches that studied 4D deep learning. Examples include application to functional magnetic resonance imaging \cite{zhao2018modeling,bengs2019a} and computed tomography \cite{clark2019convolutional,van2019stacked}. This work focuses on studying the properties of deep learning-based motion estimation and the challenging problem of learning from high-dimensional 4D spatio-temporal data. First, we design a 4D convolutional neural network (CNN) that takes an entire sequence of volumes as the input. Second, we propose a mixed 3D-4D CNN architecture for more efficient processing, that performs spatial 3D processing first, followed by full 4D processing. Third, we also make use of temporal information at the  model output by introducing a regularization strategy that forces the model to predict motion states for previous time steps within the 4D sequence. For comparison, we consider a deep learning approach using a template and a moving volume as the input \cite{gessert2019two} which is common for motion estimation \cite{laves2019deep}. In contrast to previous deep learning approaches \cite{gessert2018deep,gessert2019two}, we do not use an additional marker and estimate motion for a tissue dataset. We evaluate our best performing method with respect to robustness towards image rotations and motion distortions. In summary, our contributions are  three fold. First, we provide an extensive comparison of different deep learning architectures for estimating motion from high-dimensional 4D spatio-temporal data. Second, we propose a novel architecture that significantly outperforms previous deep learning methods. Third, we propose a novel regularization strategy, demonstrating that additional temporal information is also beneficial at the model output. 

\section{Methods }
\label{Methods}

\subsection{Experimental Setup}
\begin{figure}
\centering
\begin{tikzpicture}
\node[anchor=south west,inner sep=0] (image) at (0,0) {\includegraphics[width=0.7\textwidth]{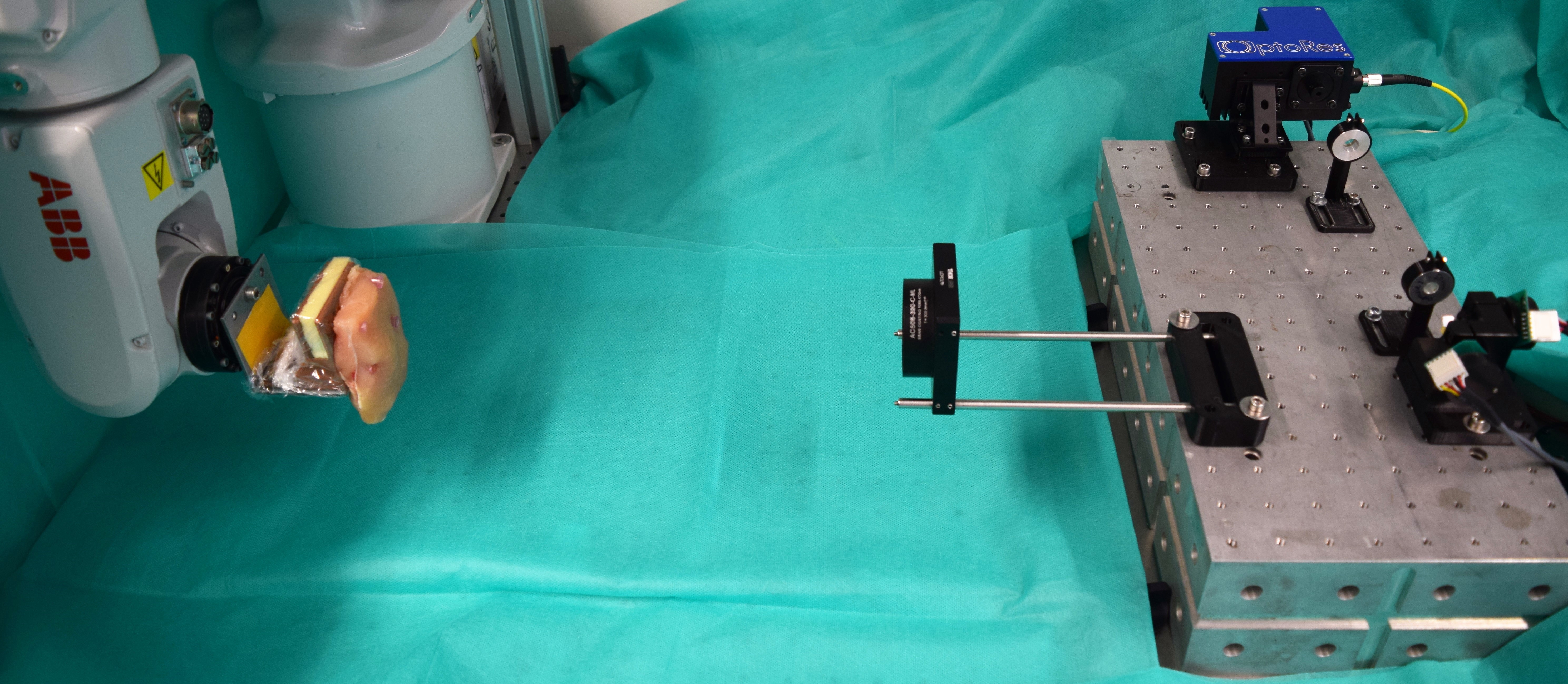}};
\node [anchor=west, white] (marker) at (0.7,0.8) {Chicken breast sample};
\node [anchor=west, white] (ABB) at (3,3) {Robot};
\node [anchor=west, white] (OCT) at (4,3) {OCT scan head};
\node [anchor=west, white] (galvos) at (6.8,0.5) {Galvos}; %(3,2)
\node [anchor=west, white] (lens) at (5,0.5) {Lens};
  
   \begin{scope}[x={(image.south east)},y={(image.north west)}]
		%\draw (1, 1) node {Hello world}        
        \draw [->, line width=1pt, white] (marker) --(0.25,0.400);
        \draw [->, line width=1pt, white] (ABB) --(0.06,0.80);
        \draw [->, line width=1pt, white] (OCT) --(0.77,0.9036);
        \draw [->, line width=1pt, white] (galvos) --(0.92,0.52);
        \draw [->, line width=1pt, white] (lens) --(0.6,0.3845);     
        \draw [->, line width=1pt, white] (lens) --(0.9,0.58); 
        \draw [->, line width=1pt, white] (lens) --(0.85,0.75); 
        
        %\draw [-latex, ultra thick, red] (note) to[out=0, in=-120] (0.48,0.80);
        %\draw [-stealth, line width=5pt, cyan] (water) -- ++(0.4,0.0);
    \end{scope}
\end{tikzpicture}%
\caption{The experimental setup for data acquisition and annotation. The chicken breast sample is attached with needles to a holder of the robot. The OCT device itself is not shown.}
\label{fig:Exp_Setup}
\end{figure}

For evaluation of our motion estimation methods, we employ a setup which allows for automatic data acquisition and annotation, see Figure~\ref{fig:Exp_Setup}. We use a commercially available swept-source OCT device (OMES, OptoRes) with a scan head, a second scanning stage with two mirror galvanometers, lenses for beam focusing and a robot (ABB IRB 120). The OCT device is able to acquire a single volume in 1.2 ms. A chicken breast sample is attached with needles to a holder of the robot. Our OCT-setup allows for shifting the FOV without moving the scan head by using the second mirror galvanometers stage and by changing the pathlength of the reference arm. Two stepper motors control the mirrors of the second scanning stage, which shift the FOV in the lateral directions. A third stepper motor changes the pathlength of the reference arm to translate the FOV in the axial dimension.   
For evaluation of our methods we consider volumes of size $32\times32\times32$  with a corresponding FOV of approximately $5\,\mathrm{mm}\times5\mathrm{\,mm}\times3.5\mathrm{\,mm}$. 

\subsection{Data Acquisition}
We consider the task of motion estimation of a given ROI with respect to its initial position. To assess our methods on various tissue regions, we consider 40 randomly chosen ROIs of a chicken breast sample with the same size as the OCT's FOV.

For motion estimation, only the relative movement between the FOV and ROI is relevant, hence moving the ROI and using a steady FOV is equivalent to moving the FOV and using a steady ROI. This can be exploited for generation of both OCT and ground-truth labels. By keeping the ROI steady and moving the FOV by a defined shift in stepper motor space, we simulate relative ROI movement. At the same time, the defined shift provides a ground-truth motion as we can transform the shift in motor space to the actual motion in image space using a hand-eye calibration.

\begin{figure}
\centering
\includegraphics[width=1.0\textwidth]{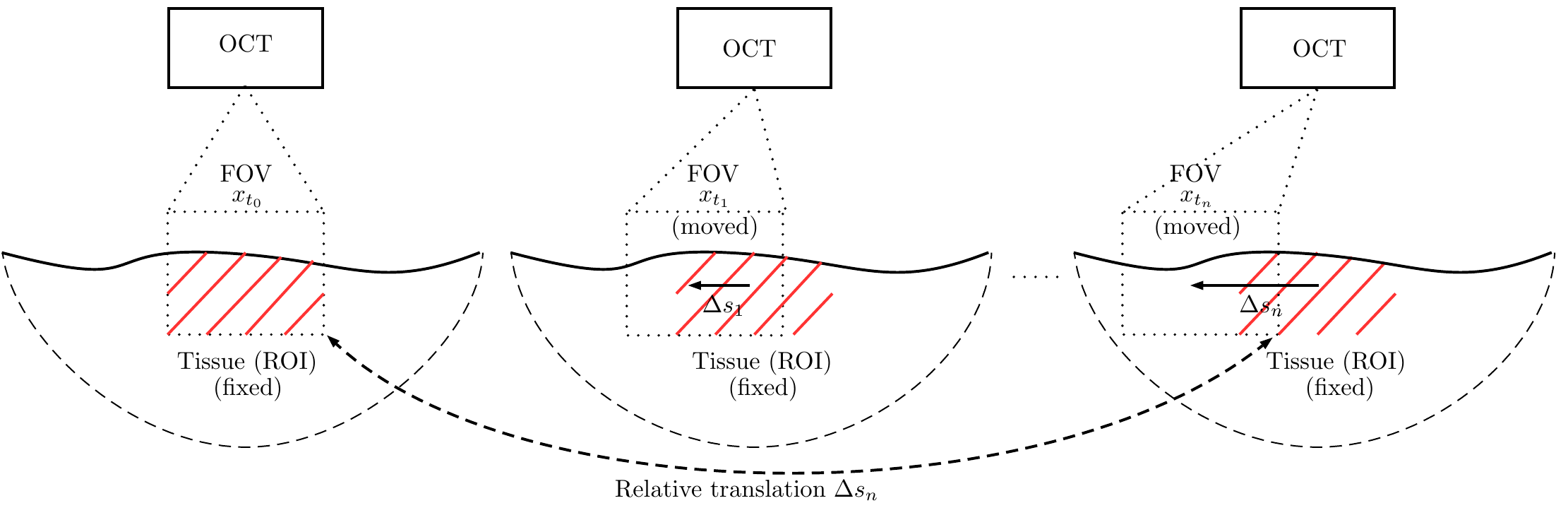}
\caption{Our data acquisition strategy. For motion estimation only the relative movement is relevant, hence we use a fixed ROI and move the FOV step-wise by $\Delta s_{i}-\Delta s_{i-1}$. This results in a sequence of OCT volumes $x_{t}$ with the corresponding relative translation $\Delta s$ between the initial volume $x_{t_{0}}$ and the last volume $x_{t_{n}}$ of a sequence.} 
\label{fig:Data_Acq}
\end{figure}

Initially, the FOV completely overlaps with the target ROI. After acquiring an initial template image volume $x_{t_{0}}$ of the ROI, we use the stepper motors to translate the FOV by $\Delta s_{t_{1}}$ such that the target ROI only partially overlaps with the FOV. Now, we acquire an image volume $x_{t_{1}}$ for the corresponding translation $\Delta s_{t_{1}}$. This step can be repeated multiple times, resulting in a sequence of shifted volumes  $x_{t_{i}}$ and known relative translations  $\Delta s_{t_{i}}$ between the initial ROI and a translated one. Note, each translation  $\Delta s_{t_{i}}$  is relative to the initial position of a ROI. The procedure is illustrated in Figure~\ref{fig:Data_Acq}. 

In this way, we formulate a supervised learning problem where we try to learn the relative translation $\Delta s_{t_{n}}$ of an ROI experiencing motion with respect to its initial position, given a sequence of volumes $x_{{t}}= \{x_{t_{0}},...,x_{t_{n}}\}$. 

\begin{figure}
\centering
\includegraphics[width=0.9\textwidth]{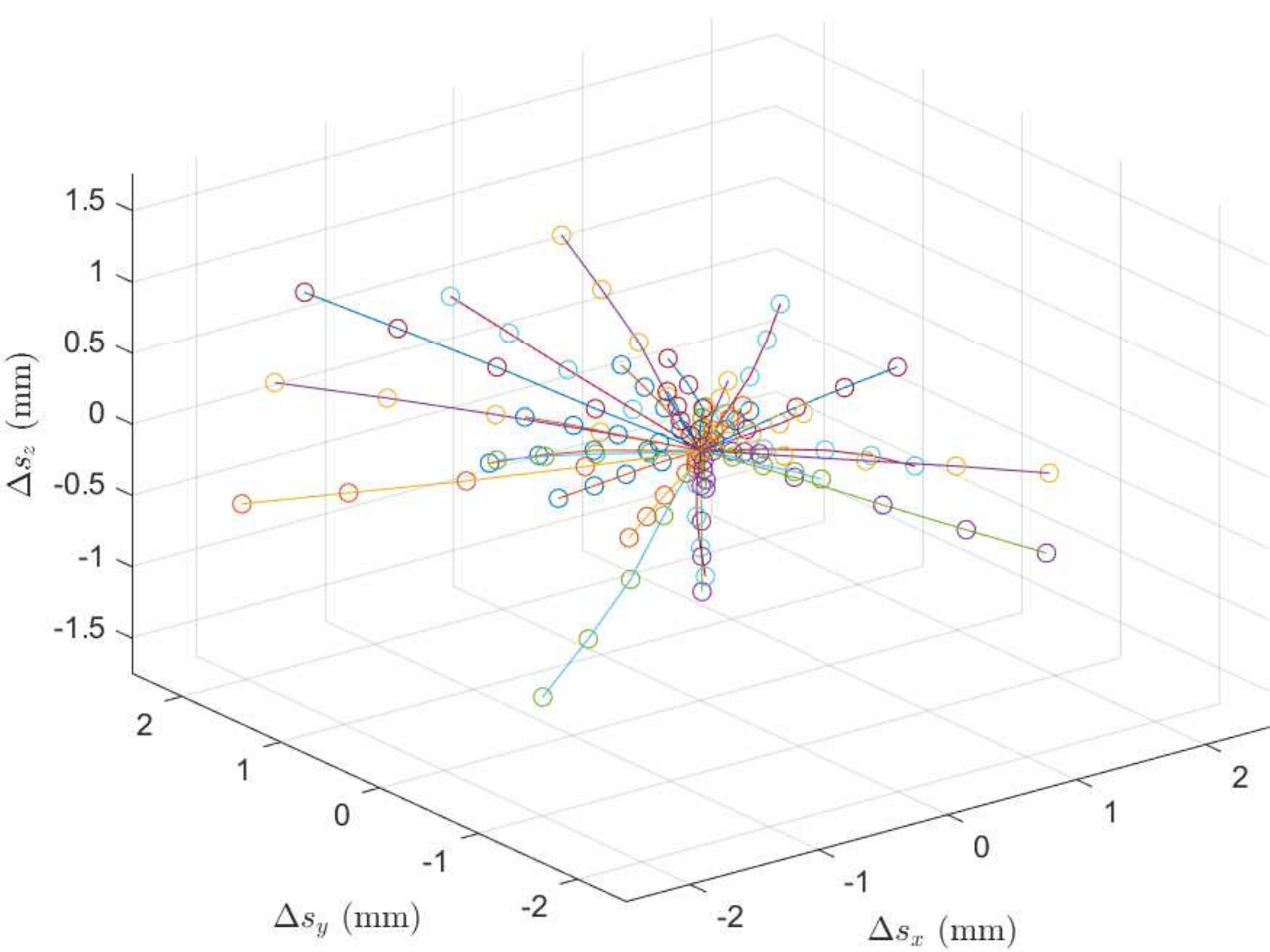}
\caption{Shown are 30 example trajectories for the translations in the spatial dimensions, each trajectory consists of a sequence of five target shifts $\Delta s_{i}$ (circle).}
\label{fig:trajectories}
\end{figure}

For generation of a single motion trajectory, we consider a sequence of five target translations, i.e., target motor shifts $\Delta s_{t} = [\Delta s_{t_{0}},\Delta s_{t_{1}},\Delta s_{t_{2}},\Delta s_{t_{3}},\Delta s_{t_{4}} ]$. To generate a smooth motion pattern, we randomly generate $\Delta s_{t_{4}}$ and use spline interpolation between $\Delta s_{t_{0}}=[0,0,0]$, $\Delta s_{t_{4}}$ and a randomly generated connection point $\Delta s_{c}$. We sample the intermediate target shifts $\Delta s_{t_{1}},\Delta s_{t_{2}},\Delta s_{t_{3}}$ from the spline function. This results in various patterns where the FOV drifts away from the ROI. By using different distances between $\Delta s_{0}$ and $\Delta s_{4}$ we simulate different magnitudes of motions and obtain various different motor shift distances between subsequent volumes. Example trajectories are shown in Figure~\ref{fig:trajectories}. We use a simple calibration between galvo motor steps and image coordinates, to transform the shifts from stepper motor space to image space, resulting in a shift in millimeters. 

For data acquisition we use the three following steps. First, we use the the robot for randomly choosing an ROI. Then, the initial state of the three motors corresponds to an FOV completely overlapping with the ROI. Second, we randomly generate a sequence of five target motor states, as described above, which shifts the FOV out of the ROI. Third, at each of the target motor states, an OCT volume is acquired. 

Overall, for each ROI, we acquire OCT volumes of 200 motion patterns, where each movement consists of five target translations and five OCT volumes. 

Moreover, we evaluate how the estimation performance is affected by relative rotations between volumes of a sequence. Note, our current scanning setup is designed for translational motion as rotation is difficult to perform using galvo mirrors. Therefore, we add rotations in a post-processing step, by rotating acquired volumes of a sequence $x_{{t}}$ around the axial axis. We define a maximal rotation $\alpha_{{max}}$  and  transform each volume of a sequence with $\widetilde{x}_{{i}}=R(\alpha_{{i}})x_{{i}}$ while $\alpha_{i}=\frac{\alpha_{max}}{4}\cdot i,\,\,\forall i\in[0,4]$. Note, $R(\alpha_{{i}})$ is the rotation matrix for rotations around the depth axis. First, we consider rotations as noise that is applied to the image data. Second, we incorporate the rotation into our motion and adapt the ground truth with respect to the rotation.

Last, we also consider the effect of fast and irregular motion, such as high frequency tremors that may cause distortion within an image. This effect is unlikely to occur with our current setup as our high acquisition frequency prevents common motion artifacts \cite{zawadzki2007correction}. Nevertheless, we perform experiments with simulated motion artifacts due to relevance for slower OCT systems. We follow the findings of previous works \cite{xu2012alignment,kraus2012motion,zawadzki2007correction} and consider motion distortions as lateral and axial shifts between B-scans of an OCT volume that has been acquired without motion distortions. In this way, we are able to augment our data with defined motion distortions in a post-processing step. To simulate different intensities of motion distortions we introduce a factor $p_{dist}$ that defines the probability that a B-scan is shifted. Also, we compare shifting the B-scans one or two pixels randomly along the spatial dimensions.

\subsection{Deep Learning Models}
\label{Deep}
\begin{figure}
\centering
\includegraphics[width=1.0\textwidth]{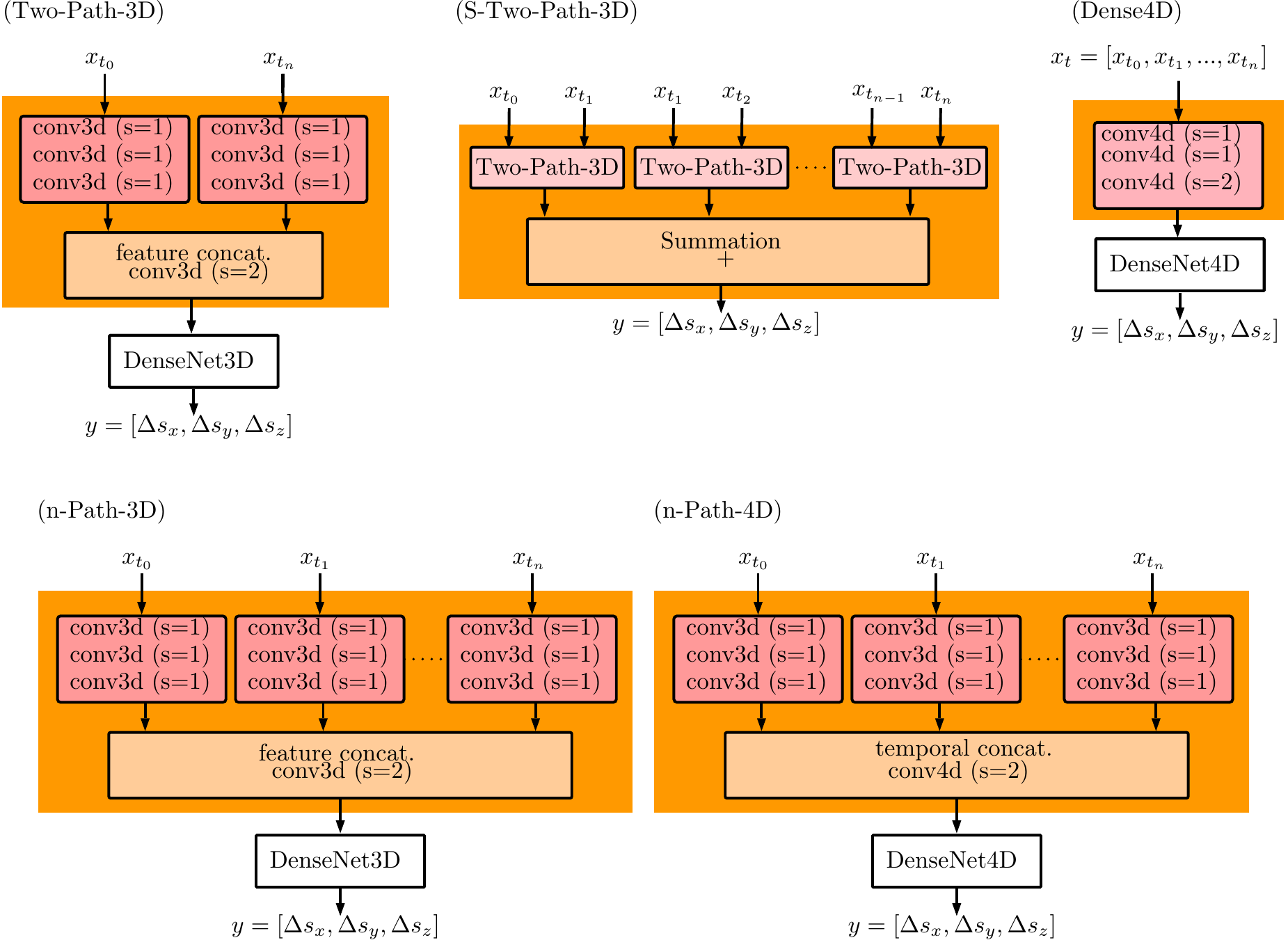}
\caption{Our proposed network architectures. The networks receive volumes $x_{t_{i}}$ from a stream of volumes to predict the motion between the volumes $x_{t_{0}}$ and  $x_{t_{n}}$. Note, for the multi path architectures, the weights are shared across the paths. }
\label{fig:methods_models}
\end{figure}

All our deep learning architectures consist of an initial processing block and a baseline block. For the baseline block we adapt the idea of densely connected neural networks (densenet) \cite{huang2017densely}. Our baseline block consists of three densenet blocks connected by average pooling layers. Each densenet block consists of 2 layers with a growth rate of 10. After the final densenet block we use a global average pooling layer (GAP) for connecting the three dimensional linear  regression output layer. Note, the output $y$ of the architecture are the relative translations between volume $x_{t_{0}}$ and  $x_{t_{n}}$ in all spatial directions. Using this baseline block, we evaluate five different initial processing concepts for motion estimation based on 4D OCT data, shown in Figure~\ref{fig:methods_models}. 
\\ \\
First, we follow the idea of a two path architecture for OCT-based motion estimation \cite{gessert2019two}. This architecture individually processes two OCT volumes up to a concatenation point by a two-path CNN with shared weights. At the concatenation point the outputs of the two paths are stacked into the channel dimension and subsequently processed jointly by a 3D CNN architecture. In this work, we use three CNN layers for the initial two-path part and our densenet baseline block with 3D convolutions (DensNet3D) for processing after the concatenation point. In the first instance we only consider the initial volume $x_{t_{0}}$ and the last volume $x_{t_{n}}$ of a sequence to estimate the relative translation. We refer to this architecture as Two-Path-3D.\\ 

Second, we use Two-Path-3D and consider predicting the relative translation between the initial and last volume, based on the sum of the relative translations between two subsequent volumes of a sequence. In this way, the network obtains information from the entire sequence. The network receives the input pairs [$x_{t_{0}}$, $x_{t_{1}}$],  [$x_{t_{1}}$, $x_{t_{2}}$],  [$x_{t_{1}}$, $x_{t_{2}}$],  [$x_{t_{2}}$, $x_{t_{3}}$],  [$x_{t_{3}}$, $x_{t_{4}}$] and the estimations are added to obtain the final network prediction $y$. Note, we train our network end-to-end based on the relative translation between the initial and the last volume and the network prediction $y$. We refer to this architecture as S-Two-Path-3D.\\

Third, we extent the idea of a two-path architecture to processing of an entire sequence of volumes, instead of using only two volumes as the networks input. For this purpose, we extend the two-path architecture to a multi-path architecture, while the number of paths is equal to number of volumes used. Note, similar to the two-path CNN, the multi-path layers consists of three layers with shared weights, followed by our densenet baseline block with 3D convolutions (DensNet3D). We refer to this architecture as Five-Path-3D.\\

Fourth, we use a 4D convolutional neural network, which employs 4D spatio-temporal convolutions and hence jointly learns features from the temporal and spatial dimensions. The input of this network is four dimensional, (three spatial and one temporal dimension) using a sequence of volumes. This method consists of an initial convolutonal part with three layers, followed by our densenet block using 4D convolutions throughout the entire network. We refer to this architecture as Dense4D.\\

Fifth, we combine the idea of 4D spatio-temporal CNNs and multi-path architectures. At first, we split the input sequence and use a multi-path 3D CNN to individually process each volume of the sequence. However, instead of concatenating the volumes along the feature dimension at the output of the multi-path CNN, we reassemble the temporal dimension by concatenating the outputs into a temporal dimension. Then, we employ our DenseNet4D baseline block. We refer to this architecture as Five-Path-4D.  

\subsection{Training and Evaluation}
We train our models to estimate the relative motion of an ROI using OCT volumes. Hence, we minimize the mean squarred error (MSE) loss function between the defined target motions $\Delta s_{t_{n}}$ and our predicted motions $y_{t_{n}}$. 

\begin{equation}
\mathcal{L}=\frac{1}{N}\sum_{j=1}^{N}\left\Vert \Delta s_{t_{n}}^{\{j\}}-y_{{t_{n}}}^{\{j\}}\right\Vert ^{2}
\end{equation}

Our goal is to estimate the relative motion between an initial volume $x_{t_{0}}$ and a final volume $x_{t_{n}}$, corresponding to the target shift $\Delta s_{t_{n}}$. Given the nature of our acquisition setup, the intermediate shifts $\Delta s_{t_{i}}$ are also available. As these additional shifts represent additional motion information, we hypothesize that they could improve model training by enforcing more consistent estimates and thus regularize the problem.

We incorporate the additional motion information by forcing our models to also predict the relative shifts of previous volumes $x_{{t_{n-1}}}$ and $x_{{t_{n-2}}}$. Thus, we also consider the relative translations $\Delta s_{t_{n-1}}$ and $\Delta s_{t_{n-2}}$ and we extent the network output by also predicting $y_{{t_{n-1}}}$ and $y_{{t_{n-2}}}$. Note, the additional output $y_{{t_{n-1}}}$ and $y_{{t_{n-2}}}$ is only considered during training and not required for application. 

For optimization, we propose and evaluate the following loss function and introduce parameters $w_{n-1}, w_{n-2}\in[0,1]$ for weighting of the additional temporal information, introduced as a regularization term. 

\begin{equation}
\mathcal{L}=\frac{1}{N}\sum_{j=1}^{N}\left\Vert \Delta s_{t_{n}}^{\{j\}}-y_{{t_{n}}}^{\{j\}}\right\Vert ^{2}+w_{n-1}\left\Vert \Delta s_{t_{n-1}}^{\{j\}}-y_{{t_{n-1}}}^{\{j\}}\right\Vert ^{2}+w_{n-2}\left\Vert \Delta s_{t_{n-2}}^{\{j\}}-y_{{t_{n-2}}}^{\{j\}}\right\Vert ^{2}
\end{equation}

We train all our models for 150 epochs, using Adam for optimization with a batch size of 50. To evaluate our models on previously unseen tissue regions we randomly choose five independent ROIs for testing and validating each. For training we use the remaining 30 ROIs. 

\label{Dataset}

\section{Results }
\label{Results}
First, we compare the different methods and report the mean absolute error (MAE), the relative mean absolute error (rMAE) and average correlation coefficient (aCC) for our experiments in Table \ref{tab:All-networks-with metrics}. The MAE is given in mm based on the calibration between galvo motor steps and image coordinates. The rMAE is calculated by dividing the MAE by targets' standard deviation. We state the number of parameters and inference times for all models, see Table~\ref{tab:network-details}. For all experiments, we test our results for significant differences in the median of the rMAE using Wilcoxon signed-rank test with $\alpha = 0.05$ significance level. Overall, using a sequence of volumes improves performance significantly and Five-Path-4D performs best with a high aCC of $98.58$\%. Comparing Five-Path-4D to Two-Path-3D, the rMAE is reduced by a factor of approximately 2.6. Moreover, employing the two-path architecture on subsequent volumes and adding the estimations (S-Two-Path-3D) performs significantly better than directly using the initial and the last volume (Two-Path-3D) of a motion sequence. 

\begin{figure}
\centering
\includegraphics[width=1.0\textwidth]{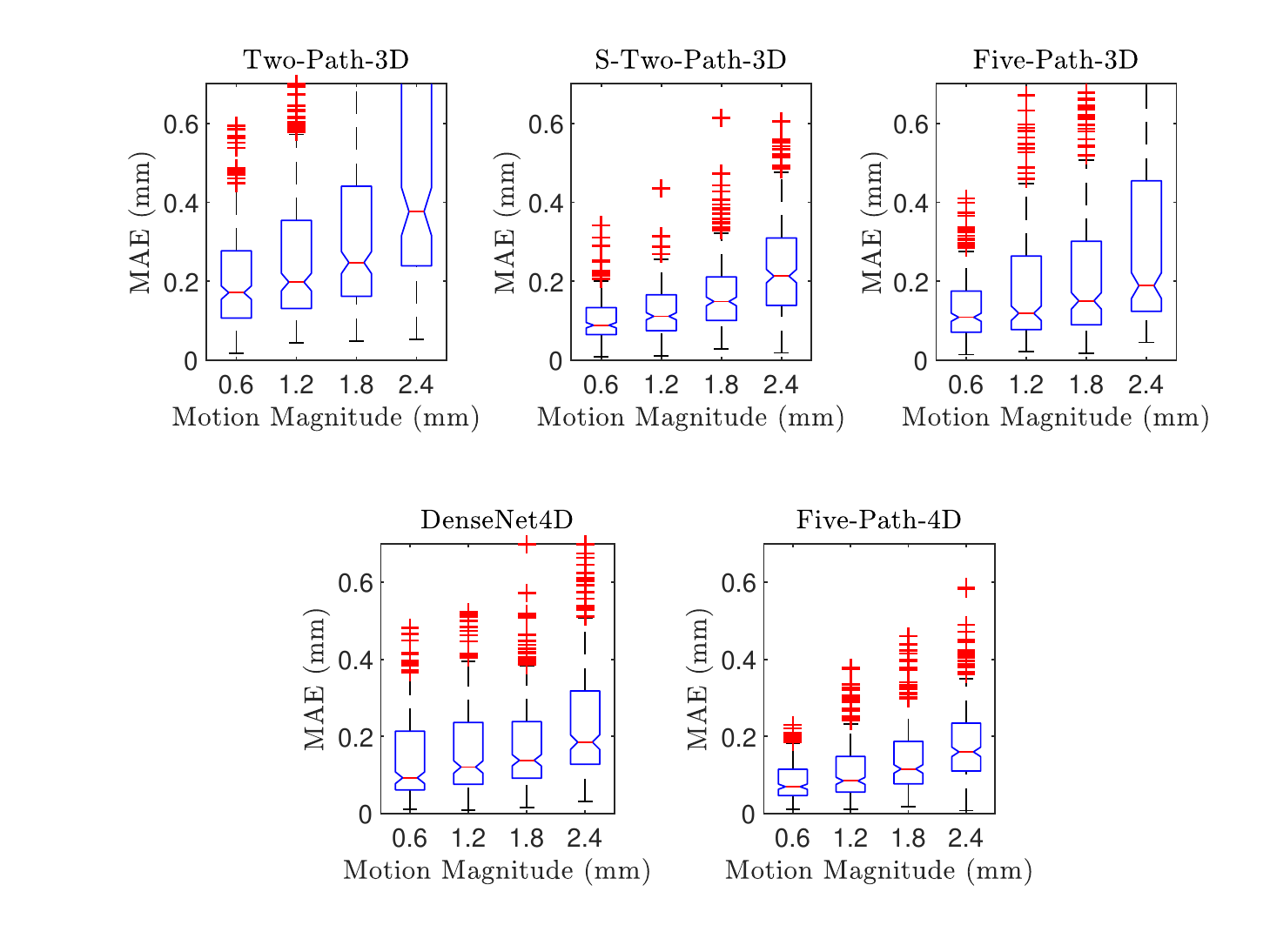}
\caption{MAE for increasing motion magnitudes. Results are shown for four motion groups, covering increasing magnitudes of motion.}
\label{fig:velocities}
\end{figure}

Second, we extent the comparison of our models and present the MAE over different motion magnitudes, shown in Figure \ref{fig:velocities}. The error increases with an increasing magnitude of the motion for all models. Comparing the different models shows that the error increases only slightly for Five-Path-4D, compared to the other models. 

Third, Table \ref{tab:Rotation_Results} shows how rotations affect the performance for our best performing model Five-Path-4D during evaluation. First, we consider rotations as noise during motion and do not transform the target shifts. Second, we consider rotations as part of the motion and transform the target shifts accordingly.  For small rotation angels $\alpha_{max}<5^{\circ}$ performance is robust and hardly reduced. For larger rotations angels $\alpha_{max}>5^{\circ}$ lateral estimation performance is affected when rotations are considered as noise, while performance remains similar when rotations are considered as part of the motion. 

%Note, the training data does not contain rotations.
Fourth, Table \ref{tab:Motion_Results} demonstrates how motion distortions affect performance. We evaluate different magnitudes of motion distortions. The results show that performance is hardly reduced when only few motion distortions are present ($p_{dist}<10\%$). However, as we increase the amount of motion distortions, performance is notably affected, yet, performance is recovered when distortions are also considered during training.

% by defining probabilities $p_{dist}$ that B-scans of an OCT volume are shifted, and also consider shifting B-scans one or two pixels
Fifth, we address the temporal regularization strategy, see Table~\ref{tab:results_different_time_steps} for our best performing model Five-Path-4D. We report performance metrics for various weighting factors $w_{n-1}$ and $w_{n-2}$. Our results demonstrate that using the regularization strategy improves performance. Fine tuning the weights improves performance significantly with a high aCC of $99.06\%$ for a weighting of $w_{n-1}=0.75$ and $w_{n-2}=0.75$.  

\begin{table}
 {\caption{Comparison of the different models for motion estimation. Our errors refer to the translation $\Delta s$ between the template and the last volume of a motion sequence. Errors are given in \si{\milli\metre}.
 %Our errors refer to the target translation $\Delta s$ between two volumes.
 } \label{tab:All-networks-with metrics}}%
\centering
  {\begin{tabular}{llllll}
  &  MAE $\Delta s_{x}$  &  MAE $\Delta s_{y}$  &  MAE $\Delta s_{z}$  & rMAE & aCC (\%)\\ \hline
  
  Two-Path-3D & $0.45\pm0.52$ & $0.42\pm0.52$ &  $0.18
\pm0.15 $ & $0.34\pm0.39$ & $85.47$ \\

S-Two-Path-3D & $0.20\pm0.21$ & $0.15\pm0.16$ &  $0.13
\pm0.12$ & $0.16\pm 0.17$ & $97.70$ \\

  Five-Path-3D & $0.35\pm 0.45$	&$0.18\pm	0.25$& $0.11\pm	0.09$	& $0.21\pm	0.26$ &	$93.39 $\\
  
  Dense4D & $0.22 \pm	0.21$ &	$0.20 \pm	0.24$ &	$0.13 \pm	0.11$	&$0.19 \pm 0.19$ &	$96.86$  \\

  \textbf{Five-Path-4D}  & $\mathbf{0.16\pm	0.18}$ &	$\mathbf{0.13\pm	0.15}$ &	$\mathbf{0.10	\pm0.09}$	 &$\mathbf{0.13\pm	0.14}$ &	$\mathbf{98.58}$  \\
  
\hline
  \end{tabular}}
\end{table}

\begin{table}
 {\caption{Number of parameters and inference times for all models.} \label{tab:network-details}}%
\centering
  {\begin{tabular}{llllll}
  &  Number of Parameters & Inf. Time \\ \hline
  Two-Path-3D & \num{143913} & $3.74\pm0.52$ ms\\
  S-Two-Path-3D & \num{143913} & $5.84\pm0.32$ ms\\
  Five-Path-3D &  \num{208713} & $5.23\pm0.27$ ms\\
  Dense4D  & \num{270283} & $9.78\pm0.74$ ms\\
 Five-Path-4D  &  \num{258323} & $9.34\pm0.67$ ms\\
  
\hline
  \end{tabular}}
\end{table}

\begin{table}
 {\caption{Evaluation of the performance for different rotation angels during motion. We evaluate the rotation as noise or as part of the motion, where the ground truth $\Delta s$ is rotated accordingly. The rotation angle $\alpha_{max}$ refers to the relative rotation between the initial template volume and the last volume of a sequence. Results are shown for the architecture Five-Path-4D. The errors refer to the translation $\Delta s$ between the template and the last volume of a motion sequence and are given in \si{\milli\metre}.
 } \label{tab:Rotation_Results}}%
\centering
  {\begin{tabular}{lllllll}
   & $\alpha_{max}$ &  MAE $\Delta s_{x}$  &  MAE $\Delta s_{y}$  &  MAE $\Delta s_{z}$  & rMAE & aCC (\%)\\ \hline

   \parbox[t]{2mm}{\multirow{4}{*}{\rotatebox[origin=c]{90}{Noise}}} & $2^\circ$  &$0.17	\pm 0.18	$& $0.13 \pm	0.15	$& $0.10 \pm	0.09 $ &	$0.13\pm	0.14	$ &$ 98.56$  \\
    &$5^\circ$  &$ 0.19 \pm	0.18 $&$ 	0.15	\pm0.15 $&$ 	0.09\pm	0.09 $&$ 0.14	\pm0.14	$&$ 98.44 $ \\
    &$10^\circ$ & $0.23	\pm0.19$&$  	0.16\pm	0.16	$&$ 0.10\pm	0.09	$&$ 0.17\pm	0.14	$&$ 97.95$  \\
    &$20^\circ$  & $0.34	\pm0.25	$&$ 0.23\pm	0.20$&$ 	0.10	\pm0.09	$&$ 0.22	\pm0.18	$&$ 96.04$  \\ 
 \hline 
   \parbox[t]{2mm}{\multirow{4}{*}{\rotatebox[origin=c]{90}{Motion}}} &$2^\circ$  & $0.16\pm	0.18	$&$ 0.13\pm	0.15	$&$ 0.09\pm	0.09	$&$ 0.13\pm	0.14$&$ 	98.60$  \\
    &$5^\circ$  & $0.16\pm	0.18	$&$ 0.14\pm	0.15	$&$ 0.10\pm	0.09	$&$ 0.14\pm	0.14$&$ 	98.55$  \\
    &$10^\circ$ & $0.17\pm	0.18	$&$ 0.16\pm	0.15	$&$ 0.10\pm	0.09	$&$ 0.15\pm	0.14$&$ 	98.35$ \\
    & $20^\circ$  &$0.19\pm	0.20	$&$ 0.23\pm	0.19	$&$ 0.10\pm	0.09	$&$ 0.18\pm	0.16$&$ 	97.48$ \\ 
\hline
  \end{tabular}}
\end{table}

\begin{table}
 {\caption{Results for Five-Path-4D when motion distortions are applied during evaluation; $p_{dist}$ refers to the probability that a B-scan is shifted. We evaluate shifting the B-scans one (E-1) or two pixels (E-2) during evaluation. Also, we consider motion distortions of two pixels during training and evaluation (T/E-2). Our errors refer to the translation $\Delta s$ between the template and the last volume of a motion sequence. Errors are given in \si{\milli\metre}}.
  \label{tab:Motion_Results}}%
\centering
  {\begin{tabular}{lllllll}
    Type &  $p_{dist}$  &  MAE $\Delta s_{x}$  &  MAE $\Delta s_{y}$  &  MAE $\Delta s_{z}$  & rMAE & aCC (\%)\\ \hline

E-1 &50\%  &$0.31\pm	0.33$&$	0.29\pm	0.29$&$	0.14\pm	0.11$&$	0.25\pm	0.24$&$	94.41$ \\
E-1 &25\% &$0.20\pm	0.22$&$	0.20\pm	0.20$&$	0.11\pm	0.10$&$	0.17\pm	0.17$&$	97.37$ \\
E-1 &10\% &$0.16\pm	0.18$&$	0.16\pm	0.17$&$	0.10\pm	0.09$&$	0.14\pm	0.15$&$	98.25$ \\
\\
E-2 &50\%  &$0.33\pm	0.35$&$	0.28\pm	0.28$&$	0.14\pm	0.12$&$	0.25\pm	0.24$&$	94.27$ \\
E-2 &25\%  & $0.20\pm	0.21$&$	0.20\pm	0.21$&$	0.12\pm	0.10$&$	0.17\pm	0.17$&$	97.39$ \\
E-2 &10\%  & $0.17\pm	0.18$&$	0.15\pm	0.16$&$	0.10\pm	0.09$&$	0.14\pm	0.14$&$	98.27$
 \\
\\
T/E-2 &50\%  &   $0.18\pm	0.21$&$	0.15\pm	0.15$&$		0.10\pm	0.08$&$		0.14\pm	0.15$&$97.97$ \\

\hline
  \end{tabular}}
\end{table}

\begin{table}
 {\caption{Evaluation of the temporal loss regularization using different weighing factors $w_{n-1}$, $w_{n-2}$. Results are shown for the architecture Five-Path-4D with respect to predicting the motion $\Delta s$ between the template and the last volume of a sequence. Errors are given in \si{\milli\metre}.}
 
 \label{tab:results_different_time_steps}}%
\centering
  {\begin{tabular}{lllllll}
  
  $w_{n-1}$ & $w_{n-2}$ &   MAE $\Delta s_{x}$  &  MAE $\Delta s_{x}$  &  MAE $\Delta s_{x}$  & rMAE & aCC (\%) \\ \hline
  
     0 & 0 &  $0.16\pm	0.18$ &	$0.13\pm	0.15$ &	$0.10	\pm0.09$	 &$0.13\pm	0.14$ &	$98.58$ \\
  
     1 & 0 & $0.15 \pm	0.22$ &	$0.12\pm	0.13$	&$0.11\pm	0.10$	& $0.13\pm	0.15$	& $98.15$  \\
  
   0.75 & 0 &  $0.14 \pm	0.13$ &	$0.11\pm	0.10$&	$0.13\pm0.10$	&$0.14	\pm0.11$& 	$98.90$ 
 \\
  
    0.5 & 0  & $0.10	\pm0.09$ &	$0.14\pm	0.11$&	$0.10\pm	0.08$&	$0.12\pm	0.10$&	$99.02$
\\
    0.25 & 0  &$ 0.11\pm	0.11$ &	$0.14\pm	0.13$&	$0.11 \pm	0.09$	&$0.12\pm	0.11$&	$98.92$

\\
  
  \\
   1 & 1 & $0.11\pm0.10$&	$0.19\pm	0.17$&	$0.10\pm	0.09$	&$0.14	\pm0.12$&	$98.71$  \\
  
   \textbf{0.75} & \textbf{0.75} & $\mathbf{0.09\pm	0.09}$&	$\mathbf{0.11\pm	0.10}$&	$\mathbf{0.10\pm	0.08}$&	$\mathbf{0.10\pm	0.09}$&	$\mathbf{99.06}$ \\

   0.75 & 0.5 &  $0.12\pm	0.10$ &	$0.10 \pm	0.11$ &	$0.10\pm	0.08$ &	$0.11\pm	0.10$& 	$99.03$
  \\

\hline
  \end{tabular}}
\label{tab:temporal_regularization}
\end{table}

\section{Discussion }
\label{Discussion}

Motion estimation is a relevant problem for intraoperative OCT applications, for example in the context of motion compensation \cite{irsch2018motion} and surgical tool navigation \cite{zhang2014optical}. While previous approaches for motion estimation relied on a template and moving images, we learn a motion vector from an entire sequences of OCT volumes. This leads to the challenging problem of 4D spatio-temporal deep learning.

We design three new CNN models that address 4D spatio-temporal processing in different ways. While Five-Path-3D is an immediate extension of the previous two-path approach \cite{gessert2019two}, our Five-Path-4D and Dense4D models perform full 4D data processing. For a fair comparison, we also consider pairwise motion estimation along the sequence using Two-Path-3D, aggregated to a final estimate. Our results in Table~\ref{tab:All-networks-with metrics} show that the two-path method using only the start and the end volume perform worse than the other methods. This demonstrates that there is not enough information for motion estimation or the motion is too large. 

For using a full sequence of volumes the Five-Path-3D CNN performs significantly worse than the other deep learning approaches. This indicates that stacking multiple volumes in the models feature channel dimension is not optimal for temporal processing. This has also been observed for spatio-temporal problems in the natural image domain \cite{tran2015learning}. This is also supported by pair-wise processing with S-Two-Path-3D which shows a significantly higher performance than the feature stacking approach and a higher performance than Dense4D. Our proposed 4D architecture outperforms all other approaches, including the previous deep learning concepts using two volumes \cite{laves2019deep,gessert2019two} and pair-wise processing. Thus, we demonstrate the effective use of full 4D spatio-temporal information with a new deep learning model.

Next, we also consider the effect of different motor shift distances for our problem. Note, faster movements lead to larger distance between subsequent volumes of a sequence and to reduced overlap, making motion estimation harder as there are fewer features for finding correspondence. The results in Figure~\ref{fig:velocities} show the performance for different distances between volumes. As expected, we observe a steady increase with larger distances for all models. For the approaches using just two volumes, the increase is substantial while it remains moderate for the 4D spatio-temporal models. Thus, 4D data is also beneficial for various magnitudes of motion to be estimated and we demonstrate that the models effectively deal with different spatial distances between time steps.

Moreover, Table \ref{tab:Rotation_Results} shows how rotations affect performance for our best performing method when applied during evaluation. When rotations are considered as noise, only for large rotations $\alpha_{max}>5^{\circ}$ performance is notably reduced. However, when rotations are considered as part of the motion, performance remains similar even for larger rotations. As rotations were not present in the training data, the results indicate that our models are robust with respect to rotations.

Furthermore, we consider the problem of potential motion artifacts. The OCT device we employ is able to acquire an OCT volume in \SI{1.2}{\milli\second}. According to Zawadzki et al., motion artifacts are not present for volume acquisition speeds below \SI{100}{\milli\second} \cite{zawadzki2007correction}. However, to ensure that our methods are applicable to slower OCT devices as well, we consider the effect of fast and irregular motion that may cause image distortions. We consider motion distortions as lateral or axial shifts between B-scans of an OCT volume, similar to previous works \cite{xu2012alignment,kraus2012motion,zawadzki2007correction}. The results in Table \ref{tab:Motion_Results} demonstrate that motion distortions applied only during evaluation can affect performance. This highlights the importance of fast volumetric imaging when 4D data is used for motion estimation. However, when motion artifacts are also considered during training, performance can be recovered. These results indicate that using deep learning with 4D data is a viable approach, even if data is affected by fast and irregular motion distortions.
% We evaluate motion distortions that shift B-scans one or two pixels. Also, we consider different probabilities that B-scans are shifted withing an OCT volume, simulating different intensities of motion distortions.

As temporal information appears to be beneficial at the model input, we also consider usage at the model output. Here, we introduce a regularization strategy which forces the model to learn consecutive motion steps. We also introduce weighting factors for fine tuning of our approach. Our results in Table~\ref{tab:results_different_time_steps} demonstrate that the regularization method appears to be effective. While a weighting equal to one does not lead to an immediate performance improvement, using a weighing of $w_{n-1}=0.75$, $w_{n-2}=0.75$ improves performance notably up to an aCC of $99.06$ \%. As a result, providing more information on the trajectory during training appears to be helpful for 4D motion estimation.

%While our 4D deep learning methods significantly improve performance, their more costly 4D convolution operations also affect inference times which is important for application when real-time processing is required. Inference times in comparison to model size are shown in Table~\ref{tab:network-details}. While two-Path-3D with pair-wise processing can provide motion estimates with up to 171 Hz, our best performing method five-Path-4D still achieves 107 Hz. Hence, our 4D deep learning methods are a viable approach for real-time motion estimation.
While our 4D deep learning methods significantly improve performance, their more costly 4D convolution operations also affect inference times which is important for application when real-time processing is required. Inference times in comparison to model size are shown in Table~\ref{tab:network-details}. While, Five-Path-4D significantly outperforms S-Two-Path-3D in terms of motion estimation performance, S-Two-Path-3D allows for faster predictions.
 Thus, there is a trade-off between performance and inference time for the different architectures. However, with an inference time of 107 Hz our 4D deep learning methods are already a viable approach for real-time motion estimation which could be improved in the future by using more powerful hardware or additional low-level software optimization.
%While two-Path-3D with pair-wise processing can provide motion estimates with up to 171 Hz, our best performing method five-Path-4D achieves 107 Hz.
\section{Conclusion }
\label{Conclusion}
We investigate deep learning methods for motion estimation using 4D spatio-temporal OCT data. We design and evaluate several 4D deep learning methods and compare them to previous approaches using a template and a moving volume. We demonstrate that our novel 3D-4D deep learning method significantly improves estimation performance on a tissue data set, compared the previous deep learning approach of using two volumes. We observe that large motion is handled well by the 4D deep learning methods. Also, we demonstrate the effectiveness of using additional temporal information at the network's output by introducing a regularization strategy that forces the 4D model to learn an extended motion pattern. These results should be considered for future applications such as motion compensation or the adjustment of surgical tools during interventions. Also, our 4D spatio-temporal methods could be extended to other problems such as ultrasound-based motion estimation.

\section*{Compliance with ethical standards}
\begin{small}
%This work was partially funded by AiF research grant number ZF4026302CR7.
\textbf{Funding:} This work was partially funded by Forschungszentrum Medizintechnik Hamburg (grants 04fmthh16). \\ 
\textbf{Conflict of interest:} The authors declare that they have no conflict of interest.\\ 
\textbf{Ethical approval:} This article does not contain any studies with human participants or animals performed by any of the authors.\\ 
\textbf{Informed consent:} Not applicable
\end{small}

% BibTeX users please use one of
%\bibliographystyle{spbasic}      % basic style, author-year citations
\bibliographystyle{spmpsci}      % mathematics and physical sciences
\bibliography{CARS_Tracking.bib}   % name your BibTeX data base

\begin{thebibliography}{10}
\providecommand{\url}[1]{{#1}}
\providecommand{\urlprefix}{URL }
\expandafter\ifx\csname urlstyle\endcsname\relax
  \providecommand{\doi}[1]{DOI~\discretionary{}{}{}#1}\else
  \providecommand{\doi}{DOI~\discretionary{}{}{}\begingroup
  \urlstyle{rm}\Url}\fi

\bibitem{bay2006surf}
Bay, H., Tuytelaars, T., Van~Gool, L.: Surf: Speeded up robust features.
\newblock In: ECCV, pp. 404--417. Springer (2006)

\bibitem{bengs2019a}
Bengs, M., Gessert, N., Schlaefer, A.: 4d spatio-temporal deep learning with 4d
  fmri data for autism spectrum disorder classification.
\newblock In: International Conference on Medical Imaging with Deep Learning
  (2019)

\bibitem{bergmeier2017workflow}
Bergmeier, J., Fitzpatrick, J.M., Daentzer, D., Majdani, O., Ortmaier, T.,
  Kahrs, L.A.: Workflow and simulation of image-to-physical registration of
  holes inside spongy bone.
\newblock International Journal of Computer Assisted Radiology and Surgery
  \textbf{12}(8), 1425--1437 (2017)

\bibitem{clark2019convolutional}
Clark, D., Badea, C.: Convolutional regularization methods for 4d, x-ray ct
  reconstruction.
\newblock In: Medical Imaging 2019: Physics of Medical Imaging, vol. 10948, p.
  109482A. International Society for Optics and Photonics (2019)

\bibitem{du2013robustness}
Du, X., Assadi, M.Z., Jowitt, F., Brett, P.N., Henshaw, S., Dalton, J., Proops,
  D.W., Coulson, C.J., Reid, A.P.: Robustness analysis of a smart surgical
  drill for cochleostomy.
\newblock The International Journal of Medical Robotics and Computer Assisted
  Surgery \textbf{9}(1), 119--126 (2013)

\bibitem{ehlers2014integrative}
Ehlers, J.P., Srivastava, S.K., Feiler, D., Noonan, A.I., Rollins, A.M., Tao,
  Y.K.: Integrative advances for oct-guided ophthalmic surgery and
  intraoperative oct: microscope integration, surgical instrumentation, and
  heads-up display surgeon feedback.
\newblock PloS one \textbf{9}(8), e105224 (2014)

\bibitem{eilers2009navigated}
Eilers, H., Baron, S., Ortmaier, T., Heimann, B., Baier, C., Rau, T.S.,
  Leinung, M., Majdani, O.: Navigated, robot assisted drilling of a minimally
  invasive cochlear access.
\newblock In: 2009 IEEE International Conference on Mechatronics, pp. 1--6.
  IEEE (2009)

\bibitem{finke2012automatic}
Finke, M., Kantelhardt, S., Schlaefer, A., Bruder, R., Lankenau, E., Giese, A.,
  Schweikard, A.: Automatic scanning of large tissue areas in neurosurgery
  using optical coherence tomography.
\newblock The International Journal of Medical Robotics and Computer Assisted
  Surgery \textbf{8}(3), 327--336 (2012)

\bibitem{gessert2019two}
Gessert, N., Gromniak, M., Schl{\"u}ter, M., Schlaefer, A.: Two-path 3d cnns
  for calibration of system parameters for oct-based motion compensation.
\newblock In: Medical Imaging 2019: Image-Guided Procedures, Robotic
  Interventions, and Modeling, vol. 10951, p. 1095108. International Society
  for Optics and Photonics (2019)

\bibitem{gessert2018deep}
Gessert, N., Schl{\"u}ter, M., Schlaefer, A.: A deep learning approach for pose
  estimation from volumetric oct data.
\newblock Medical image analysis \textbf{46}, 162--179 (2018)

\bibitem{huang2017densely}
Huang, G., Liu, Z., Van Der~Maaten, L., Weinberger, K.Q.: Densely connected
  convolutional networks.
\newblock In: CVPR, pp. 4700--4708 (2017)

\bibitem{ilg2017flownet}
Ilg, E., Mayer, N., Saikia, T., Keuper, M., Dosovitskiy, A., Brox, T.: Flownet
  2.0: Evolution of optical flow estimation with deep networks.
\newblock In: CVPR, pp. 2462--2470 (2017)

\bibitem{irsch2018motion}
Irsch, K., Lee, S., Bose, S.N., Kang, J.U.: Motion-compensated optical
  coherence tomography using envelope-based surface detection and kalman-based
  prediction.
\newblock In: Advanced Biomedical and Clinical Diagnostic and Surgical Guidance
  Systems XVI, vol. 10484, p. 104840Q. International Society for Optics and
  Photonics (2018)

\bibitem{kraus2012motion}
Kraus, M.F., Potsaid, B., Mayer, M.A., Bock, R., Baumann, B., Liu, J.J.,
  Hornegger, J., Fujimoto, J.G.: Motion correction in optical coherence
  tomography volumes on a per a-scan basis using orthogonal scan patterns.
\newblock Biomedical optics express \textbf{3}(6), 1182--1199 (2012)

\bibitem{lankenau2007combining}
Lankenau, E., Klinger, D., Winter, C., Malik, A., M{\"u}ller, H.H., Oelckers,
  S., Pau, H.W., Just, T., H{\"u}ttmann, G.: Combining optical coherence
  tomography (oct) with an operating microscope.
\newblock In: Advances in medical engineering, pp. 343--348. Springer (2007)

\bibitem{laves2019deep}
Laves, M.H., Ihler, S., Kahrs, L.A., Ortmaier, T.: Deep-learning-based 2.5 d
  flow field estimation for maximum intensity projections of 4d optical
  coherence tomography.
\newblock In: Medical Imaging 2019: Image-Guided Procedures, Robotic
  Interventions, and Modeling, vol. 10951, p. 109510R. International Society
  for Optics and Photonics (2019)

\bibitem{laves2017feature}
Laves, M.H., Schoob, A., Kahrs, L.A., Pfeiffer, T., Huber, R., Ortmaier, T.:
  Feature tracking for automated volume of interest stabilization on 4d-oct
  images.
\newblock In: Medical Imaging 2017: Image-Guided Procedures, Robotic
  Interventions, and Modeling, vol. 10135, p. 101350W. International Society
  for Optics and Photonics (2017)

\bibitem{van2019stacked}
van~de Leemput, S.C., Prokop, M., van Ginneken, B., Manniesing, R.: Stacked
  bidirectional convolutional lstms for deriving 3d non-contrast ct from
  spatiotemporal 4d ct.
\newblock IEEE transactions on medical imaging  (2019)

\bibitem{lowe1999object}
Lowe, D.G.: Object recognition from local scale-invariant features.
\newblock In: ICCV, vol.~99, pp. 1150--1157 (1999)

\bibitem{pau2008imaging}
Pau, H., Lankenau, E., Just, T., H{\"u}ttmann, G.: Imaging of cochlear
  structures by optical coherence tomography (oct). temporal bone experiments
  for an oct-guided cochleostomy technique.
\newblock Laryngo-rhino-otologie \textbf{87}(9), 641--646 (2008)

\bibitem{schluter2019feasibility}
Schl{\"u}ter, M., Otte, C., Saathoff, T., Gessert, N., Schlaefer, A.:
  Feasibility of a markerless tracking system based on optical coherence
  tomography.
\newblock In: Medical Imaging 2019: Image-Guided Procedures, Robotic
  Interventions, and Modeling, vol. 10951, p. 1095107. International Society
  for Optics and Photonics (2019)

\bibitem{siddiqui2018high}
Siddiqui, M., Nam, A.S., Tozburun, S., Lippok, N., Blatter, C., Vakoc, B.J.:
  High-speed optical coherence tomography by circular interferometric ranging.
\newblock Nature photonics \textbf{12}(2), 111 (2018)

\bibitem{tran2015learning}
Tran, D., Bourdev, L., Fergus, R., Torresani, L., Paluri, M.: Learning
  spatiotemporal features with 3d convolutional networks.
\newblock In: ICCV, pp. 4489--4497 (2015)

\bibitem{vienola2012real}
Vienola, K.V., Braaf, B., Sheehy, C.K., Yang, Q., Tiruveedhula, P., Arathorn,
  D.W., de~Boer, J.F., Roorda, A.: Real-time eye motion compensation for oct
  imaging with tracking slo.
\newblock Biomedical optics express \textbf{3}(11), 2950--2963 (2012)

\bibitem{wang2016heartbeat}
Wang, T., Pfeiffer, T., Regar, E., Wieser, W., van Beusekom, H., Lancee, C.T.,
  Springeling, G., Krabbendam-Peters, I., van~der Steen, A.F., Huber, R., van
  Soest, G.: Heartbeat oct and motion-free 3d in vivo coronary artery
  microscopy.
\newblock JACC: Cardiovascular Imaging \textbf{9}(5), 622--623 (2016)

\bibitem{xu2012alignment}
Xu, J., Ishikawa, H., Wollstein, G., Kagemann, L., Schuman, J.S.: Alignment of
  3-d optical coherence tomography scans to correct eye movement using a
  particle filtering.
\newblock IEEE transactions on medical imaging \textbf{31}(7), 1337--1345
  (2012)

\bibitem{zawadzki2007correction}
Zawadzki, R.J., Fuller, A.R., Choi, S.S., Wiley, D.F., Hamann, B., Werner,
  J.S.: Correction of motion artifacts and scanning beam distortions in 3d
  ophthalmic optical coherence tomography imaging.
\newblock In: Ophthalmic Technologies XVII, vol. 6426, p. 642607. International
  Society for Optics and Photonics (2007)

\bibitem{zhang2014optical}
Zhang, Y., Pfeiffer, T., Weller, M., Wieser, W., Huber, R., Raczkowsky, J.,
  Schipper, J., W{\"o}rn, H., Klenzner, T.: Optical coherence tomography guided
  laser cochleostomy: Towards the accuracy on tens of micrometer scale.
\newblock BioMed research international \textbf{2014} (2014)

\bibitem{zhao2018modeling}
Zhao, Y., Li, X., Zhang, W., Zhao, S., Makkie, M., Zhang, M., Li, Q., Liu, T.:
  Modeling 4d fmri data via spatio-temporal convolutional neural networks
  (st-cnn).
\newblock In: International Conference on Medical Image Computing and
  Computer-Assisted Intervention, pp. 181--189. Springer (2018)

\end{thebibliography}

\end{document}